\documentclass{elsart}

 \usepackage{graphicx}

\usepackage{amssymb}

\begin{document}

\begin{frontmatter}



\title{Dislocations in cubic crystals described by discrete models }

\author[UC3M]{ L. L. Bonilla \corauthref{cor}},
\corauth[cor]{Corresponding author.}
\ead{bonilla@ing.uc3m.es}
\author[UCM]{A. Carpio},
\ead{ana$_{-}$carpio@mat.ucm.es}
\author[UC3M]{ I. Plans},
\ead{ignacio.plans@uc3m.es }

\address[UC3M]{Modelizaci\'on, Simulaci\'on Num\'erica y Matem\'atica
Industrial, 
Universidad Carlos III de Madrid, Avenida de la Universidad 30, 28911 Legan{\'e}s, Spain}
\address[UCM]{Departamento de Matem\'{a}tica Aplicada, Universidad
Complutense de Madrid, 28040 Madrid, Spain}

\date{ \today  }

\begin{abstract}
Discrete models of dislocations in cubic crystal lattices having one or two atoms
per unit cell are proposed. These models have the standard linear anisotropic elasticity as 
their continuum limit and their main ingredients are the elastic stiffness constants of the
material and a dimensionless periodic function that restores the translation invariance of 
the crystal and influences the dislocation size. For these models, conservative and 
damped equations of motion are proposed. In the latter case, the entropy production 
and thermodynamic forces are calculated and fluctuation terms obeying the
fluctuation-dissipation theorem are added. Numerical simulations illustrate static 
perfect screw and 60$^\circ$ dislocations for GaAs and Si. 

\end{abstract}

\end{frontmatter}

\section{Introduction}
Heteroepitaxial growth is fundamental for manufacturing important nanoelectronic devices 
based on self-assembled quantum dots \cite{kim03,zha04} or superlattices \cite{gra95}. 
Powerful imaging techniques based on scanning probe microscopes have been developed
which allow to monitor these crystal growth processes down to atomic distances. Moreover,
these techniques help visualizing defects such as misfit dislocations 
\cite{hul01,MBl74,BRB97,CHw95} (which may act as nucleation sites for quantum dots
\cite{kim03}) down to their cores. Understanding epitaxial growth requires understanding
and relating processes which cover a wide range of scales \cite{che04}. Other multiscale
problems need to be solved if we try to understand how dislocations \cite{nab67,hir82}, 
grain boundaries \cite{hul01}, cracks \cite{fre90} and other defects control the 
mechanical, optical and electronic properties of the resulting materials \cite{science}. 
Understanding these multiscale processes requires understanding better the relation between 
the defects themselves and observed macroscopic behavior, which remains an active area 
of research. While over mesoscopic length scales a coarse grained description based on
defect densities makes sense \cite{holt,vinals}, the cores of defects need to be resolved on 
nanometric scales and then coupled to the coarser description. Coarse grained descriptions
of dislocation dynamics include irreversible thermodynamics, as in the works by Holt 
\cite{holt} and by Rickman and Vi\~nals \cite{vinals}, Boltzmann-type kinetic equations, 
as in the formulations by  Groma and coworkers \cite{groma} (and references therein), or
stochastic models \cite{hahner}. 
The atomic scale near defects can be resolved by {\em ab initio} or molecular dynamics 
simulations, which are very costly at the present time. Thus, it is 
interesting to have systematic models of defect motion in crystals that can be solved cheaply, 
are compatible with elasticity and yield useful information about the defect cores and their 
mobility.

In a previous paper, we have proposed a discrete model of dislocations and their motion in 
cubic crystals with a one atom basis \cite{CB05}. In this paper, we present an extension of 
our previous theory to treat crystals with two-atom basis in their primitive cells (such as the 
diamond and zinc-blende structures of silicon and gallium arsenide, respectively). Moreover, 
we explain how to include dissipation in the dynamics of the model and how to consider the 
effect of fluctuations by using the ideas of fluctuating hydrodynamics \cite{ll6,sbm82,rr98}. 
Our model covers length scales in the nanometer range. In principle, 
to make contact with existing mesoscopic theories \cite{vinals,groma,hahner}, one should 
define a dislocation density tensor and coarse grain over length scales up to hundreds of 
nanometers. This is outside the scope of the present work.

The main ingredients entering our discrete model are the elastic stiffness constants of the 
material and a dimensionless periodic function that restores the translation invariance of the 
crystal and influences the dislocation size. To be precise, consider a simple cubic symmetry 
with one atom per lattice point. Firstly, we discretize space along the primitive vectors 
defining the unit cell of the crystal ${\bf x}\equiv (x,y,z)=(l,m,n) a$, in which $a$ is the 
length of the primitive cubic cell, and $l$, $m$ and $n$ are integer numbers. Secondly, we 
replace the gradient of the displacement vector $\tilde{u}_i(x,y,z,t)= a\, u_i(l,m,n;t)$ 
($u_i(l,m,n;t)$ is a nondimensional vector) in the strain energy density by an appropriate 
periodic function of the discrete gradient, $g(D_{j}^+ u_{i})$: We shall define the {\em 
discrete} distortion tensor as
\begin{eqnarray} 
&& w_{i}^{(j)} = g(D^+_j u_i), \label{e1}\\
&& D^\pm_1 u_i(l,m,n;t) = \pm\, [u_i(l\pm 1,m,n;t) - u_i(l,m,n;t)], \label{e2}
\end{eqnarray} 
etc., where $g(x)$ is a periodic function of period one satisfying $g(x)\sim x$ as $x\to 0$. 
The strain energy density for the discrete model is obtained by substituting the strain tensor 
in the usual strain energy density:
\begin{eqnarray} 
&&  W = {1\over 2} c_{ijkl} e_{ij} e_{kl}, \label{e3}\\
&& c_{ijkl} = C_{12}\, \delta_{ij} \delta_{kl} + {C_{11}-C_{12}\over 2}\,
(\delta_{ik} \delta_{jl} + \delta_{il} \delta_{jk})  \nonumber \\ 
&& \, + H\,\left( {\delta_{ik} \delta_{jl} + \delta_{il} 
\delta_{jk}\over 2} -\delta_{1i} \delta_{1j}\delta_{1k} \delta_{1l}
  - \delta_{2i} \delta_{2j}\delta_{2k} \delta_{2l} -\delta_{3i} \delta_{3j}
\delta_{3k} \delta_{3l} \right) , \quad\label{e4}\\
&& H = 2C_{44}+C_{12}-C_{11}, \label{e5}\\ 
&& e_{ij} = {1\over 2}\, (w_{i}^{(j)} + w_{j}^{(i)}) = {g(D^+_j u_i) +
g(D^+_i u_j)\over 2} \label{e6} 
\end{eqnarray} 
(sum over repeated indices is assumed). Here $\lambda = C_{12}$, $\mu = (C_{11} - 
C_{12})/2$ are the usual Lam\'e coefficients if $H= 0$ and therefore the crystal is isotropic. 
Summing over all lattice sites, we obtain the potential energy of the crystal:
\begin{eqnarray} 
V(\{u_i\}) = a^3\, \sum_{l,m,n} W(l,m,n;t) , \label{e7}
\end{eqnarray}
in which we have considered the strain energy deensity to be a function of the point 
$W({\bf u})=W(l,m,n;t)$, $(l,m,n)=(x,y,z)/a$. Next, we find the equations of motion 
with or without dissipation by the usual methods of classical mechanics. For conservative 
dynamics:
\begin{eqnarray} 
\rho a^4\, \ddot{u}_{i}(l,m,n;t) = -{1\over a}\, 
{\partial V(\{u_k\})\over \partial u_i(l,m,n;t) } , \label{e8}
\end{eqnarray} 
or, equivalently (see Section \ref{sec:model})\cite{CB05},
\begin{eqnarray} 
\rho a^2\, \ddot{u}_{i} = \sum_{j,k,l} D^-_j [c_{ijkl}\, g'(D^+_j u_i)\, 
g(D^+_l u_k)],     \label{e9}
\end{eqnarray}
Here $\ddot{u}_{i} \equiv \partial^2u_{i}/\partial t^2$ and the displacement vector 
is dimensionless, so that both sides of Eq.\ (\ref{e9}) have units of force per unit area. Let 
us now restore dimensional units to Equation (\ref{e9}), so that $\tilde{u}_i(x,y,z)= a\, 
u_i(x/a,y/a,z/a)$, then let $a\to 0$, use Eq.\ (\ref{e9}) and that $g(x)\sim x$ as $x\to 0$. 
Then we obtain the usual Cauchy equations of linear elasticity:
\begin{eqnarray} 
\rho\, {\partial^2\tilde{u}_{i}\over\partial t^2} &=& \sum_{j,k,l} {\partial
\over\partial x_j}\left(c_{ijkl}\, {\partial \tilde{u}_k\over\partial x_l}\right), 
\label{e11}
\end{eqnarray}
provided the components of the distortion tensor are very small. Far from the core of a 
defect, the discrete gradient approaches the continuous one. Then, provided the slope $g'(0)$ 
is one in the appropriate units, the spatially discrete equations of motion become those of the 
anisotropic elasticity. 

The periodic function $g(x)$ ensures that sliding a plane of atoms an integer number of 
times the lattice distance $a$ parallel to a primitive direction does not change the potential
energy of the crystal. We choose
\begin{equation} 
g(x) = \left\{ \begin{array} {ll}
x , \quad |x|< {1\over 2} - \alpha,\\
{(1 -2\alpha) (1 - 2x)\over 4\alpha}\,,
\quad {1\over 2} - \alpha < x < {1\over 2} + \alpha, \end{array} \right. 
\label{e12}
\end{equation} 
which is periodically extended outside the interval $(\alpha-1/2,\alpha+1/2)$ for a given 
$\alpha\in (0,1/2)$. To select $\alpha$, we calculate the Peierls stress needed to move
a given dislocation as a function of $\alpha$ and fit it to data from experiments or
molecular dynamics calculations. Once the discrete model is specified, different dislocation 
configurations can be selected by requiring that their far field should adopt the well-known 
form of continuous elasticity \cite{CB05}.

The rest of the paper is organized as follows. In Section \ref{sec:model}, we review the
derivation of the governing equations with conservative dynamics for simple cubic symmetry,
and give the numerical constructions of screw and edge dislocations. We use 
the well known screw and edge dislocations for anisotropic elasticity to set up the boundary 
conditions far from the dislocation core and the initial conditions in overdamped equations 
of motion. Numerical solution of these equations yields the static dislocation configuration 
of our discrete elasticity model. In Section \ref{dissipation} we include dissipation 
and fluctuations in the equations of motion. Dissipation is described by a Rayleigh dissipative
function that is a quadratic functional of the strain rate tensor, which, in turn, depends on the
discrete distorsion tensor. Since the distortion tensor (containing finite differences of the
displacement vector) and its rate are larger near the core of defects, we expect that dissipation 
will be stronger near the core of a moving dislocation than at its far field. Fluctuations are 
introduced via the fluctuation-dissipation theorem and they should be stronger near the core 
of moving dislocations. An extension of our ideas to crystals with more complicated 
symmetries requires formulating our equations in non-orthogonal coordinates, which is 
explained in Section \ref{sec:fcc}. The equations of motion for two-atom bases are obtained 
in Section \ref{sec:2basis} and the corresponding screw and 60$^\circ$ perfect dislocations 
are calculated for diamond and zinc-blende structures. Section \ref{sec:conclusions} 
contains our conclusions. 

\section{Conservative equations of motion for a simple cubic lattice} 
\label{sec:model}
In this Section, we shall derive the equations of motion (\ref{e9}) for the conservative
dynamics given by (\ref{e8}). Firstly, let us notice that 
\begin{eqnarray} 
{\partial W\over \partial u_i(l,m,n;t)} = {\partial W\over\partial e_{jk}}
{\partial e_{jk}\over \partial u_i(l,m,n;t)} = {1\over 2}\, \sigma_{jk}\,
{\partial[g(D^+_j u_k) + g(D^+_k u_j)]\over \partial u_i(l,m,n;t)}\nonumber\\
 = {1\over 2}\, \sigma_{jk}\,\left[ g'(D^+_j u_k) {\partial (D^+_j u_k)
\over \partial u_i(l,m,n;t)}+ g'(D^+_k u_j) {\partial (D^+_k u_j)\over \partial
u_i(l,m,n;t)}\right] , \label{B1}
\end{eqnarray} 
where $W$ is a function of the point $(l',m',n')$, and we have used the
definition of stress tensor:
\begin{eqnarray}
\sigma_{ij} = {\partial W\over\partial e_{ij}} , \label{B2}
\end{eqnarray} 
and its symmetry, $\sigma_{ij} = \sigma_{ji}$. Now, we have
\begin{eqnarray} 
{\partial \over \partial u_i(l,m,n;t) }[D^+_1 u_k(l',m',n';t)] =\delta_{ik}
\, (\delta_{l\, l'+1} - \delta_{ll'})\, \delta_{m m'} \delta_{n n'} ,
\label{B3}
\end{eqnarray} 
and similar expressions for $j= 2,3$. By using (\ref{B1}) - (\ref{B3}), we
obtain
\begin{eqnarray} {\partial\over \partial u_i(l,m,n;t)} \sum_{l',m',n'}
W(l',m',n';t) = - \sum_j D^-_j  [\sigma_{ij}\,g'(D^+_j u_i)]
\,. \label{B4}
\end{eqnarray} 
In this expression, no sum is intended over the subscript $i$, so that we have abandoned 
the Einstein convention and explicitly included a sum over $j$. Therefore Eq.\ (\ref{e8}) 
for conservative dynamics becomes
\begin{eqnarray} 
\rho a^2\, \ddot{u}_{i} &=& \sum_j D^-_j [\sigma_{ij}\,g'(D^+_j u_i)],
\label{B5}
\end{eqnarray}
which yields Eq.\ (\ref{e9}). Except for the factor $g'(D^+_j u_i)$, these equations 
are discretized versions of the usual ones in elasticity \cite{ll7}.

\subsection{Static dislocations of the discrete model}
To find the dislocation solutions of our model, we need the stationary solution of the 
anisotropic elasticity equations at zero applied stress corresponding to the same type of 
dislocation. In all cases, the procedure to obtain numerically the dislocation from
the discrete model is the same. We first solve the stationary equations of elasticity with
appropriate singular source terms to obtain the {\em dimensional} displacement vector 
${\bf \tilde{u}}(x,y,z) = (\tilde{u}_{1}(x,y,z),\tilde{u}_{2}(x,y,z),\tilde{u}_{3}
(x,y,z))$ of the static dislocation {\em under zero applied stress}. This displacement vector 
yields the far field of the corresponding dislocation for the discrete model, which is the {\em 
nondimensional} displacement vector: 
\begin{eqnarray}
{\bf U}(l,m,n)= \frac{{\bf \tilde{u}}\left((l+\delta_{1})a,(m+\delta_{2})a,(n+
\delta_{3})a\right)}{a} .    \label{em1}
\end{eqnarray}
Here $0\leq\delta_{i}<1$, $i=1,2,3$, are chosen so that the singularity at $x=y=z=0$ does 
not coincide with a lattice point. For a sc crystal, it is often convenient to select the center of 
a unit cell, $\delta_{i}=1/2$. We use the nondimensional static displacement vector ${\bf 
U}(l,m,n)$ defined by (\ref{em1}) in the boundary and initial conditions for the discrete 
equations of motion. 

Take for example, a pure screw dislocation along the $z$ axis with Burgers vector ${\bf b}= 
(0,0,\textrm{b})$ has a displacement vector ${\bf \tilde{u}} =(0,0,\tilde{u}_{3}(x,y))$ 
with $\tilde{u}_{3}(x,y)= \textrm{b}\, (2\pi)^{-1} \tan^{-1} (y/x)$ \cite{nab67}. 
The discrete equation for the $z$ component of the {\em nondimensional} displacement 
$u_{3}(l,m;t)$ is:
\begin{eqnarray}
\rho a^2\, \ddot{u}_{3} &=& C_{44}\, \{D^-_1 [g(D^+_1 u_{3})\, g'(D^+_1 u_{3})] 
+ D^-_2 [g(D^+_2 u_{3})\, g'(D^+_2 u_{3})]\}. \label{em2}
\end{eqnarray}
Numerical solutions of Eq.\ (\ref{em2}) show that a static screw dislocation moves if 
an applied shear stress surpasses the static Peierls stress, $|F|<F_{cs}$, but that a moving 
dislocation continues doing so until the applied shear stress falls below a lower threshold 
$F_{cd}$ (dynamic Peierls stress); see Ref.~\cite{car03} for a similar situation 
for edge dislocations. To find the static solution of this equation corresponding to a screw 
dislocation, we could minimize an energy functional. However, it is more efficient to solve 
the following overdamped equation:
\begin{eqnarray}
\beta\, \dot{u}_{3} &=& C_{44}\, \{D^-_1 [g(D^+_1 u_{3})\, g'(D^+_1 u_{3})] 
+ D^-_2 [g(D^+_2 u_{3})\, g'(D^+_2 u_{3})]\}.   \label{em3}
\end{eqnarray}
The stationary solutions of Eqs.\ (\ref{em2}) and (\ref{em3}) are the same, but the 
solutions of (\ref{em3}) relax rapidly to the stationary solutions if we choose appropriately 
the damping coefficient $\beta$. We solve Eq.\ (\ref{em3}) with initial condition 
$u_{3}(l,m;0)= U_{3}(l,m)\equiv \textrm{b}\, (2\pi a)^{-1} \tan^{-1}[(m+1/2)/(l+1/2)]$
(corresponding to $\delta_{i}=1/2$), and with boundary conditions $u_{3}(l,m;t)= 
U_{3}(l,m)+ F\, m$ at the upper and lower boundaries of our lattice. At the lateral 
boundaries, we use zero-flux Neumann boundary conditions. Here $F$ is an applied 
dimensionless stress with $|F|<F_{cs}$ (the dimensional stress is $C_{44}F$). For this small 
stress, the solution of Eq.\ (\ref{em3}) relaxes to a static screw dislocation $u_{3}(l,m)$ 
with the desired far field. Figure 3 of Ref.~\cite{CB05} shows the helical structure adopted 
by the deformed lattice $(l,m,n+u_{3}(l,m))$ for an asymmetric piecewise linear $g(x)$ as in 
Eq.\ (\ref{e11}). The numerical solution shows that moving a dislocation requires that we 
should have $g'(D_{j}^+u_{3})<0$ (with either $j=1$ or 2) at its core \cite{car03}, which 
is harder to achieve as $\alpha$ decreases. A discusion of the changes in the 
size of the dislocation core and the Peierls stress due to $\alpha$ can be found in 
Ref.~\cite{CB05}; see in particular Figure 2. Using the same technique,
stationary planar edge dislocations for an isotropic sc material have 
been constructed and a variety of dipole and loops of edge 
dislocations have been numerically found \cite{CB05}. 

\section{Dissipative equations of motion and fluctuations}
\label{dissipation}
\subsection{Equations of motion including dissipation}
Overdamped dynamics obtained by replacing the time differential of the displacement
vector instead of the inertial term in the equation of motion (\ref{e9}) is not too realistic. 
Instead, we can add dissipation to the equations of motion by considering a quadratic
dissipative function with cubic symmetry:
\begin{eqnarray} 
R &=&  \left(\zeta -{2\over 3}\eta\right) {\dot{e}_{ll}^2\over 2} +
\eta \dot{e}_{ik}^2 + {\gamma\over 2} (\dot{e}_{ik}- \dot{e}_{11}
\delta_{1i} \delta_{1k} - \dot{e}_{22} \delta_{2i} \delta_{2k} -\dot{e}_{33}
\delta_{3i} \delta_{3k})^2\quad\quad  \label{B6}
\end{eqnarray} 
For an isotropic body, we have $\gamma=0$ and then $\zeta$ and $\eta$ are the
usual viscosities; see Eq.\ (34.5) in Ref.~\cite{ll7}. The viscous part
of the stress tensor is the symmetric tensor
\begin{eqnarray}
&&\Sigma_{ik} = {\partial R\over\partial \dot{e}_{ik}} =
\eta_{iklm} \dot{e}_{lm},   \label{em5}\\
&&\eta_{iklm}= {1\over 2} \left(\zeta -{2\over 3}\eta\right) \delta_{ik}
\delta_{lm} + {\eta\over 2}\, (\delta_{il}\delta_{km} + \delta_{im} 
\delta_{kl})   \nonumber \\ 
&& \, + {\gamma\over 2} \left( {\delta_{il} \delta_{km} + \delta_{im}
\delta_{kl}\over 2} -\delta_{1i} \delta_{1k} \delta_{1l}\delta_{1m}  -
\delta_{2i} \delta_{2k} \delta_{2l} \delta_{2m} -\delta_{3i} \delta_{3k}
\delta_{3l} \delta_{3m} \right).     \label{em6}
\end{eqnarray} 
In the cubic case, the viscosity tensor $\eta_{iklm}$ is determined by the
three scalar quantities $\zeta$, $\eta$ and $\gamma$. For isotropic sc 
crystals, $\gamma=0$. Similarly to Eq.\ (\ref{B4}), we can show that
\begin{eqnarray} 
{\partial\over \partial \dot{u}_i(l,m,n;t)} \sum_{l',m',n'} R(l',m',n';t) = -
\sum_j D^-_j  [\Sigma_{ij}\,g'(D^+_j u_i)]\,,\label{B9}
\end{eqnarray} 
is minus the dissipative force acting on $u_i$. Then the equation of motion
including dissipation becomes
\begin{eqnarray} 
\rho a^2\, \ddot{u}_{i} &=& \sum_j D^-_j [(\sigma_{ij}+ \Sigma_{ij})\, g'(D^+_j
u_i)].  \label{em4}
\end{eqnarray}
In the isotropic case and taking the continuum limit $a\to 0$, Eqs.\ (\ref{em4}) with 
(\ref{em5}) and (\ref{em6}) yield the viscous Navier's equations for isotropic elasticity 
\cite{ll7}:
\begin{eqnarray} 
\rho\, {\partial^2{\bf \tilde{u}}\over\partial t^2} = \mu\, \Delta {\bf\tilde{
u}} + (\lambda + \mu)\, \nabla (\nabla\cdot {\bf \tilde{u}}) + \eta\,\Delta 
{\partial {\bf \tilde{u}}\over\partial t} + \left(\zeta + {\eta\over 3}\right)\, 
\nabla \left(\nabla\cdot {\partial {\bf \tilde{u}}\over\partial t}\right). 
\label{em7}
\end{eqnarray} 

\subsection{Entropy production and fluctuations}
Fluctuations may be included in our formulation by using the ideas of Fluctuating
Hydrodynamics \cite{ll6,sbm82}. We need to find the entropy production and
write it as sum of generalized forces and fluxes. Then both the forces and the fluxes
are identified. The linear relations between forces and fluxes then yield the correlations of 
the fluctuating quantities to be added to the equations of motion.

To find the production of entropy, we need to derive a few formulas. Multiplying the 
conservative equations of motion for the model (\ref{B5}) by $\dot{u}_i$ and summing, 
we obtain 
\begin{eqnarray} 
{d\over dt}\, \sum_{l,m,n}\left[\sum_i {\rho a^2\over 2}\dot{u}^2_{i} + W(l,m,n;t)
\right] = 0 , \label{B11}
\end{eqnarray} 
after some algebra. Provided viscous terms are included in the equation of
motion, as in Eq.\ (\ref{em4}), we find 
\begin{eqnarray} 
{d\over dt}\, \sum_{l,m,n}\left[\sum_i {\rho a^2\over 2}\dot{u}^2_{i} + 
W(l,m,n;t) \right] = \sum_{l,m,n} \sum_{i,j} \dot{u}_i\, D^-_j
[\Sigma_{ij}\, g'(D^+_j u_i)]. \label{B12}
\end{eqnarray} 
Let $D^-_j \phi = \phi - \phi^-$, where $\phi= \phi(l,m,n)$ is a function of
the point $(l,m,n)$. (Then $\phi^- = \phi(l-1,m,n)$ for $j=1$, and so on). We
have $\psi D^-_j\phi = D^-_j( \psi\phi) - \phi^- D^-_j\psi$, and
\begin{eqnarray}
\sum_{l,m,n} \psi D^-_j\phi &=& \sum_{l,m,n} D^-_j( \psi\phi) -
\sum_{l,m,n}\phi^- D^-_j\psi \nonumber\\
&=& \sum_{l,m,n} D^-_j( \psi\phi) -
\sum_{l,m,n}\phi D^+_j\psi,\label{B13}
\end{eqnarray} 
after a trivial relabelling of indices. Using this formula, Eq.\ (\ref{B12})
becomes
\begin{eqnarray}
 \sum_{l,m,n} \left\{{d\over dt}\,\left[\sum_i {\rho a^2\over 2}\dot{u}^2_{i} +
 W(l,m,n;t) \right] - \sum_{i,j} D^-_j [\dot{u}_i\,\Sigma_{ij}
\, g'(D^+_j u_i)]\right\} \nonumber\\ 
= - \sum_{l,m,n} \sum_{i,j} \Sigma_{ij}\, g'(D^+_j u_i)\, (D^+_j\dot{u}_i) =
- \sum_{l,m,n} \sum_{i,j} \Sigma_{ij}\,
\dot{e}_{ij},  \label{B14}
\end{eqnarray} 
where the symmetry of the viscous stress tensor has been used to derive the
last equality. Eq.\ (\ref{B14}) describes the production of internal energy due to 
viscous processes. 

If the temperature is not homogeneous, we need to replace the strain energy density to leading
order by the elastic Helmholtz free energy density: 
\begin{eqnarray}
F({\bf u};T) = F_{0}(T) - (T-T_{0}) \alpha_{ij}e_{ij} + {1\over 2}\,
c_{ijkl}e_{ij}e_{kl}, \label{B15}
\end{eqnarray} 
in which the symmetric tensor $\alpha_{ij}$ describes anisotropic thermal expansion and 
sum over repeated indices is again implied \cite{ll7}. Here the material is undeformed at 
temperature $T_{0}$ in the absence of external forces and we assume that the temperature 
change $(T-T_{0})$ which accompanies thermoelastic deformation is small (linear 
thermoelasticity). The stress tensor is now 
\begin{eqnarray}
\sigma_{ij}=  c_{ijkl}e_{kl} - \alpha_{ij} (T-T_{0}), \label{B16}
\end{eqnarray} 
which should be inserted in the equations of motion (\ref{B5}) or (\ref{em4}). The 
entropy density is $S({\bf u};T)= -dF_{0}/dT + \alpha_{ij}e_{ij}$ if we ignore the 
temperature dependence of the elastic constants. Heat conduction is governed by the equation
$T\,\partial S/\partial t= - \partial q_{i}/\partial x_{i}$, i.e.,
\begin{eqnarray}
\rho c {\partial T\over\partial t}+ \alpha_{ij}T {\partial e_{ij}\over\partial t} 
= - {\partial q_i\over\partial x_i},\quad\quad q_i = - \kappa_{ij}\, {\partial T
\over \partial x_{j}}. \label{B17}
\end{eqnarray} 
Here $c$ is the specific heat of the solid and $\kappa_{ij}$ is the symmetric thermal 
conductivity tensor. These equations become
\begin{eqnarray}
\rho a c\, \dot{T} + a\,\alpha_{ij}T\, g'(D_{j}^+u_{i})\, D^+_{j}\dot{u}_{i}= - 
D^-_i Q_i, \quad\quad Q_i = - \kappa_{ij} {D^+_j T\over a} , \label{B18} 
\end{eqnarray} 
after discretizing. Eq.\ (\ref{B14}) can be rewritten as
\begin{eqnarray}
 &&\sum_{l,m,n} \left\{{d\over dt}\,\left[\sum_i {\rho a^2\over 2}\dot{u}^2_{i} 
 + W(l,m,n;t) \right] - \sum_{j} D^-_j \left[\sum_i \dot{u}_i\Sigma_{ij}
g'(D^+_j u_i) -  {Q_j\over a}\right]\right\} \nonumber\\  
&& \quad = - \sum_{l,m,n}\sum_{j} \left(\sum_i \Sigma_{ij}\,\dot{e}_{ij} - 
{D^-_j Q_j\over a}\right). \label{B19}
\end{eqnarray} 
The right side of this equation is related to the specific entropy (entropy per unit mass) 
$s$ by
\begin{eqnarray} 
\rho a T {\partial s\over\partial t} =  a\, \sum_{ij}
\Sigma_{ij}\,\dot{e}_{ij} - \sum_j D^-_j Q_j.  \label{B20}
\end{eqnarray} 
This can be written as
\begin{eqnarray} 
\rho {\partial s\over\partial t} =  \sum_{ij}\Sigma_{ij}\,{\dot{e}_{ij}
\over T} - \sum_j {1\over aT}\, D^-_j Q_j = \sum_{ij}\Sigma_{ij}\,{\dot{e}_{ij}
\over T} + \sum_j Q^-_j D^-_j {1\over aT} - \sum_j D^-_j {Q_j\over aT}, 
\nonumber
\end{eqnarray} 
and summing over all points, we find the entropy production:
\begin{eqnarray}
a^3\,\sum_{l,m,n} \left[ {\partial (\rho s)\over\partial t} + \sum_j D^-_j {Q_j
\over aT}\right] = \sum_{l,m,n}\left[ \sum_{ij} a^3\Sigma_{ij}\, {\dot{e}_{ij}
\over T} + \sum_j Q_j D^+_j {a^2\over T} \right] . \label{B21}
\end{eqnarray} 
This means that the generalized forces associated to the generalized velocities
$\Sigma_{ij}$ and $Q_j$ are $-a^3\dot{e}_{ij}/T$ and $- a^2\, D^+_j (1/T)$,
respectively. Eqs.\ $\Sigma_{ij} = \eta_{ijlm} \dot{e}_{lm}$ and $Q_i = (T^2/a)
\kappa_{ij} D^+_j(1/T)$ then imply that the kinetic coefficients associated to
$\Sigma_{ij}$ and $Q_j$ are $k_{B}T \eta_{ijlm}/a^3$ and $k_{B}T^2 \kappa_{ij}/a^3$,
respectively. Following Onsager's ideas as used in Fluctuating Hydrodynamics 
\cite{ll6,sbm82,rr98}, we conclude that the equations of motion including thermoelastic 
effects, dissipation and zero-mean fluctuations are as follows
\begin{eqnarray} 
&& \rho a^2\, \ddot{u}_{i} = \sum_j D^-_j [(\sigma_{ij}+ \Sigma_{ij} +
s_{ij})\,g'(D^+_j u_i)], \label{em8}\\
&& \langle s_{ij}\rangle=0, \nonumber\\
&& \langle s_{ij}(l,m,n;t) s_{ab}(l',m',n';t') \rangle = k_{B}T {\eta_{ijab} +
\eta_{abij}\over a^3}\delta_{ll'} \delta_{mm'} \delta_{nn'} \delta(t-t'),
\quad \quad \quad\label{em9}\\ 
&& \rho c\, \dot{T} + T\,\sum_{ij}\alpha_{ij} g'(D^+_{j}u_{i})\, D^+_{j}
\dot{u}_{i} = - {1\over a}\,\sum_{i} D^-_i (Q_i + \xi_i), \label{em10}\\
&& \langle \xi_i\rangle = 0,\nonumber\\
&& \langle \xi_i(l,m,n;t) \xi_j(l',m',n';t')\rangle = k_{B}T^2\, {\kappa_{ij} +
\kappa_{ji}\over a^3}\, \delta_{ll'} \delta_{mm'} \delta_{nn'}\, \delta(t-t'),
\label{em11}
\end{eqnarray}
with $\sigma_{ij}$ given by (\ref{B16}). In principle, fluctuations can be included 
in boundary conditions by using the nonequilibrium fluctuating hydrodynamics formalism
as explained in \cite{BAM} and in \cite{gomila} for the case of semiconductor interfaces.
 In crystals 
with cubic symmetry, the elastic constants and the viscosity tensor are given by Eqs.\ 
(\ref{e4}) and (\ref{em6}), respectively. The thermal conductivity and thermal 
expansion tensors are isotropic, $\kappa_{ij} = \kappa \delta_{ij}$, $\alpha_{ij} = 
\alpha \delta_{ij}$. Note that the correlations of $s_{ij}$ in (\ref{em9}) and of 
$\xi_{i}$ in (\ref{em11}) are proportional to $1/a^3$, which becomes $\delta({\bf x}
-{\bf x'})$ in the continuum limit as $a\to 0$.

Note that in our model, dissipation and fluctuations affect all atoms of the cubic lattice 
although we would expect from physical considerations that dissipation and fluctuations 
should be more pronounced near the core of moving dislocations, as they are directly related
to the motion of the atomic constituents in the core vicinity. However our model
should also fulfill these expectations. Why? Dissipation is described by a Rayleigh dissipative 
function that is a quadratic functional of the strain rate tensor, which, in turn, depends on the 
discrete distorsion tensor. Since the distortion tensor (containing finite differences of the 
displacement vector) and its rate are larger near the core of defects, we expect that dissipation 
will be stronger near the core of a moving dislocation than at its far field. Fluctuations are 
introduced via the fluctuation-dissipation theorem and they should also be stronger near the 
core of moving dislocations. 

\section{Models of fcc and bcc crystals with one atom per lattice site}
\label{sec:fcc}
In this Section, we explain how to extend our discrete models of dislocations to fcc or bcc
crystal symmetry, assuming that we have one atom per lattice site \cite{CB05}. For fcc 
or bcc crystals, the primitive vectors of the unit cell are not orthogonal. To find a 
discrete model for these crystals, we should start by writing the strain energy density in a 
non-orthogonal vector basis, $a_{1}$, $a_{2}$, $a_{3}$, instead of the usual orthonormal 
vector basis $e_1$, $e_2$, $e_3$ determined by the cube sides. Let $x_i$ denote coordinates 
in the basis $e_i$, and let $x'_i$ denote coordinates in the basis $a_i$. Notice that the $x_i$ 
have dimensions of length while the $x'_i$ are dimensionless. The matrix $\mathcal{T}=
(a_1,a_2,a_3)$ whose columns are the coordinates of the new basis vectors in terms of the 
old orthonormal basis can be used to change coordinates as follows:
\begin{eqnarray}
x_i'=\mathcal{T}_{ij}^{-1}x_j, \; x_i=\mathcal{T}_{ij}x'_j . \label{fcc2}
\end{eqnarray}
Similarly, the displacement vectors in both basis are related by
\begin{eqnarray}
u_i'=\mathcal{T}_{ij}^{-1}\tilde{u}_j, \; \tilde{u}_i=\mathcal{T}_{ij}u'_j , 
\label{fcc3}
\end{eqnarray}
and partial derivatives obey
\begin{eqnarray}
{\partial \over \partial x_i'}=\mathcal{T}_{ji} {\partial \over \partial x_j},
 \;{\partial \over \partial x_i}=\mathcal{T}_{ji}^{-1} {\partial \over \partial x_j'} .
\label{fcc4}
\end{eqnarray}
Note that $u'_{i}$ and $x'_{i}$ are nondimensional while $\tilde{u}_{i}$ and $x_{i}$ 
have dimensions of length. By using these equations, the strain energy density $W= (1/2) 
c_{iklm} e_{ik} e_{lm}$ can be written as
\begin{eqnarray}
W= {1\over 2}\, c_{ijlm}{\partial \tilde{u}_i \over \partial x_j}{\partial 
\tilde{u}_l \over\partial x_m}= {1\over 2}\, c_{rspq}' {\partial u_r'\over 
\partial x_s'} {\partial u_p' \over \partial x_q'}, \label{fcc5}
\end{eqnarray}
where the new elastic constants are:
\begin{eqnarray}
c_{rspq}'=c_{ijlm}\mathcal{T}_{ir}\mathcal{T}_{sj}^{-1}\mathcal{T}_{lp}
\mathcal{T}_{qm}^{-1}. \label{fcc6}
\end{eqnarray}
Notice that the elastic constants have the same dimensions in both the orthogonal and the 
non-orthogonal basis. To obtain a discrete model, we shall consider that the dimensionless
displacement vector $u'_{i}$ depends on dimensionless coordinates $x'_{i}$ that are integer
numbers $u'_{i}= u'_{i}(l,m,n;t)$. As in Section \ref{sec:model}, we replace the distortion 
tensor (gradient of the displacement vector in the non-orthogonal basis) by a periodic 
function of the corresponding forward difference, $w_{i}^{(j)} = g(D^+_j u'_i)$. As in 
Eq.\ (\ref{e12}),  $g$ is a periodic function with $g'(0)=1$ and period $1$. The discretized 
strain energy density is
\begin{eqnarray} 
W(l,m,n;t) = {1\over 2 }\, c_{rspq}' g(D^+_s u_r')\, g(D^+_q u_p'). \label{fcc7}
\end{eqnarray}
The elastic constants $c_{rspq}'$ in (\ref{fcc6}) can be calculated in terms of the Voigt 
stiffness constants for a cubic crystal, $C_{11}$, $C_{44}$ and $C_{12}$, which determine 
the tensor of elastic constants (\ref{e4}). The elastic energy can be obtained from Eq.\
(\ref{fcc7}) for $W$ by means of Eqs.\ (\ref{e7}). Then the conservative equations of 
motion (\ref{e8}) are
\begin{eqnarray}
\rho a^3\, {\partial^2 u'_i \over \partial t^2}= -\mathcal{T}^{-1}_{iq}
\mathcal{T}_{pq}^{-1} {\partial V \over \partial u_p'}, \nonumber 
\end{eqnarray}
which, together with Eqs.\ (\ref{e7}) and (\ref{fcc7}), yield
\begin{eqnarray}
\rho\, {\partial^2 u'_i \over \partial t^2}= \mathcal{T}^{-1}_{iq} \mathcal{T}_{
pq}^{-1}\, D^-_j [g'(D^+_ju_p')\, c_{pjrs}'\, g(D^+_s u_r')] . \label{fcc8}
\end{eqnarray}
This equation becomes (\ref{e9}) for orthogonal coordinates, $\mathcal{T}^{-1}_{iq}=
\delta_{iq}/a$.

To add dissipation and fluctuations to these equations, we need to replace $c_{pjrs}' g(D^+_s 
u_r')$ by $c_{pjrs}' g(D^+_s u_r')-\alpha'_{pj}(T-T_{0})+\eta_{pjrs}' g'(D^+_s u_r')\, 
D^+_s\dot{u}_r'+s'_{pj}$, in which $\eta_{pjrs}'$ is related to the viscosity tensor 
(\ref{em6}) in the same way as $c_{pjrs}'$ is related to $c_{ijlm}$ by (\ref{fcc6}). The 
random stress tensor $s'_{pj}$ has zero mean and correlation given by (\ref{em9}) with the 
modified viscosity tensor $\eta'_{ijab}$ instead of the viscosity tensor (\ref{em6}). The 
heat conduction equations are
\begin{eqnarray} 
&& \rho c\,{\partial T\over\partial t}  + T\,\alpha'_{ij} g'(D_{j}^+u'_{i})
\, D_{j}^+{\partial u'_i \over \partial t} = D^-_i \left(\kappa'_{ij}D^+_{j}T + {
\xi'_i\over a}\right), \label{fcc9}\\
&& \langle \xi'_i\rangle = 0,\nonumber\\
&& \langle \xi'_i(l,m,n;t) \xi'_j(l',m',n';t')\rangle 
= k_{B}T^2\, {\kappa'_{ij} + \kappa'_{ji}\over a}\, \delta_{ll'} \delta_{mm'} 
\delta_{nn'}\, \delta(t-t'),     \label{fcc10}\\
&&\kappa'_{pq} = \mathcal{T}_{pi}^{-1}\mathcal{T}_{qj}^{-1}\kappa_{ij}, \quad 
\alpha'_{pq} = {1\over 2}\left(\mathcal{T}_{ip}\mathcal{T}_{qj}^{-1} +
\mathcal{T}_{jp}\mathcal{T}_{qi}^{-1}\right)\alpha_{ij}.\label{em12}
\end{eqnarray}
Note that the both the original and the modified tensors $\alpha_{ij}$ and $\kappa_{ij}$ 
are symmetric.

Once we have derived the equations of motion, stationary dislocations can be calculated by
first finding the corresponding solution to the equations of anisotropic elasticity and using
it to set up initial and boundary conditions for overdamped equations of motion. For fcc
and bcc crystals, screw and edge dislocations have been constructed in Ref.~\cite{CB05}.

\section{Models for diamond and zincblende structures}
\label{sec:2basis}
Silicon and gallium arsenide are semiconductors of great importance for industry that
crystalize in the face centered cubic (fcc) system. Crystals of these materials can be 
described as a fcc Bravais lattice with a basis of two atoms per site, which constitute 
a diamond structure for Si and a zincblende structure for GaAs \cite{grahn}.
When growing layers of these materials, defects are very important because they act 
as nucleation sites, and have to be eliminated after the growth process has ceased. Among 
defects, dislocations and misfit dislocations are often observed \cite{MBl74,hul01}. 
Thus, it is desirable to have an economic description of these defects and their dynamics in 
terms of control parameters such as temperature \cite{science}. A molecular dynamics 
description is very costly if we need to couple atomic details in the nanoscale to a mesoscopic 
description in larger scales that are important in the growth process \cite{caflisch}. In this 
Section, we extend the previous models for cubic crystals with an atom per lattice site to 
crystals having two atoms per site (extension to crystals with more than two atoms per site is 
straightforward). Having two or more atoms per site introduces new features that 
are better explained revisiting the classic Born-von Karman work on vibrations of a linear 
diatomic chainÊ\cite{BK}. We shall show how to obtain the wave equation for acoustic 
phonons in the elastic limit, directly from the equations for the diatomic chain. A similar 
calculation allows us in Subsection \ref{sec:model2} to obtain the 
Cauchy equations for anisotropic elasticity in the continuum limit of our discrete models, 
which are constructed with the aim of having exactly this property. Subsection \ref{sec:gaas}
shows how to calculate static dislocations for GaAs and Si.

\subsection{Continuum limit for the linear diatomic chain}
\label{sec:BK}
We shall consider a diatomic chain comprising alternatively atoms of masses $M_{1}$ and 
$M_{2}$ whose equilibrium positions are separated a distance $a/2$. The atoms are restricted
to move only along the length of the chain. Their displacement with respect to their 
equilibrium positions will be denoted by $a\, u_{l}$ and $a\, v_{l}$, respectively, in which 
$l$ is the cell index, and $u_{l}$ and $v_{l}$ are dimensionless. If $\phi$ is the quadratic 
potential of interaction between neighboring atoms, the equations of motion for the diatomic 
chain are \cite{BK}
\begin{eqnarray}
M_{1} \ddot{u}_{l} &=& {1\over a}\left[\phi'\left((v_{l}- u_{l})\, a+ {a\over 2}
\right) - \phi'\left((u_{l}- v_{l-1})\, a+ {a\over 2}\right)\right]\nonumber\\
&=& \phi''\left({a
\over 2}\right)\, [(v_{l}- u_{l}) - (u_{l}- v_{l-1})], \label{bk1}\\
M_{2} \ddot{v}_{l} &=& {1\over a}\left[\phi'\left((u_{l+1}- v_{l})\, a+ {a\over 2}
\right) -\phi'\left((v_{l}- u_{l})\, a+ {a\over 2}\right)\right]\nonumber\\
& =& \phi''\left({a
\over 2}\right)\, [(u_{l+1}- v_{l}) - (v_{l}- u_{l})] .   \label{bk2}
\end{eqnarray}
If we assume that the solutions of these equations are plane waves,
\begin{eqnarray}
u_{l} = U\, e^{i\, (2\pi\eta l -\omega t)}, \quad\quad
v_{l} = V\, e^{i\, (2\pi\eta l -\omega t)},   \label{bk3}
\end{eqnarray}
the following dispersion relation is obtained 
\begin{eqnarray}
\omega^2 = {\phi''(a/2)\over M_{1}M_{2}}\, [(M_{1}+M_{2})
\mp \sqrt{(M_{1}+M_{2})^2- 4M_{1}M_{2}\sin^2\pi\eta}], \label{bk4}
\end{eqnarray}
in which the minus (resp., plus) sign corresponds to the acoustic (resp., optic) branch
of the dispersion relation \cite{BK}. Moreover, the corresponding amplitude ratio
for the acoustic branch is 
\begin{eqnarray}
{U\over V} = {-M_{2}\, (1+e^{-i\, 2\pi\eta})\over (M_{1}-M_{2}) - 
\sqrt{(M_{1}+M_{2})^2- 4M_{1}M_{2}\sin^2\pi\eta}},  \label{bk5}
\end{eqnarray}
with a similar formula for the optical branch \cite{BK}. In the long wavelength limit,
$\eta\to 0$, the acoustic vibrations satisfy
\begin{eqnarray}
U&=& V,\quad \omega = c\, {2\pi\eta\over a}, \label{bk6}\\
c &=& \sqrt{{\phi''(a/2)\, a^2\over 2\, (M_{1}+M_{2})}} = \sqrt{{E\over\rho}}, 
\label{bk7}\\
\rho &=& {M_{1}+M_{2}\over a}, \quad E= {\phi''(a/2)\, a\over 2}. \label{bk8}
\end{eqnarray}
In these equations, $E$ and $\rho$ are the Young modulus and the linear mass density,
respectively \cite{BK}. In the limit as $\eta\to 0$, each cell comprising two atoms 
moves rigidly with a phase velocity $c$ and a wave number $2\pi\eta/a$. 

The continuum limit of the diatomic chain equations recovers the acoustic vibrations only. 
In this limit, $l\to\infty$ and $a\to 0$, with fixed $x=la$. Furthermore, 
\begin{eqnarray}
a\, u_{l}(t)= \tilde{u}(la,t) = \tilde{u}(x,t), \quad a\, v_{l}(t)= \tilde{u}\left( 
la + {a\over 2},t\right) = \tilde{u}\left( x + {a\over 2},t\right).  \label{bk9}
\end{eqnarray}
If we now add Eqs.\ (\ref{bk1}) and (\ref{bk2}) divide by $a$, and use (\ref{bk9}) 
to approximate the result, we obtain the following wave equation in the continuum limit:
\begin{eqnarray}
\rho\, {\partial^2\tilde{u}\over\partial t^2} = E\, {\partial^2\tilde{u}\over 
\partial x^2},    \label{bk10}
\end{eqnarray}
provided $\rho$ and $E$ are given by Eq.\ (\ref{bk8}). The wave speed $c$ is then given 
by Eq.\ (\ref{bk7}). Equation (\ref{bk10}) is the elastic continuum 
limit of the diatomic chain equations, {\em which does not contain optical vibrations}.

\subsection{Discrete model for a fcc lattice with a two-atom basis}
\label{sec:model2}
We shall now propose a discrete model for a fcc lattice with a basis comprising
two atoms, of masses $M_{1}$ and $M_{2}$, respectively. Although this model is much
more complicated to describe, the key ideas to show that it is compatible with anisotropic
elasticity are the same as in Subsection \ref{sec:BK} for the diatomic chain. 

The main ideas needed to write a model for this crystal structure are the following:
\begin{enumerate}
\item Write the strain energy corresponding to a fcc crystal in a non-orthogonal basis
with axes given by the usual primitive directions of the fcc Bravais lattice.
\item Write the corresponding strain energy for a fcc crystal with two atoms per lattice
site.
\item Restore the periodicity of the crystal by defining the discrete distortion tensor as a 
periodic function (with period 1) of the discrete gradient of the displacement vector.
\item Define the potential energy of the crystal as the strain energy times the volume of 
the unit cell summed over all lattice sites. Then write down the equations of motion for the 
displacement vectors at each site.
\item Check that the continuum limit of the model yields the usual anisotropic elasticity.
\end{enumerate}

We shall now carry out this program, which is an extension of that presented in Section
\ref{sec:fcc} for a fcc lattice with a single atom per site; see also Ref.~\cite{CB05}. 
The primitive vectors of the fcc lattice are
\begin{eqnarray}
a_1= {a\over 2}\, (0,1,1), \; \quad a_2= {a\over 2}\, (1,0,1), \; 
\quad a_3={a\over 2}\, (1,1,0),
\label{m1}
\end{eqnarray}
in terms of the usual orthonormal vector basis $e_1$, $e_2$, $e_3$ determined by the cube. 
From these vectors, we determine the matrix $\mathcal{T}_{ij}$ to change coordinates as in 
(\ref{fcc2}) and (\ref{fcc3}). In the continuum limit, the strain energy is given by 
(\ref{fcc5}) with elastic constants given by (\ref{fcc6}) and (\ref{e4}).

\begin{center}
\begin{figure}[h]
\begin{center}
\includegraphics[width=12cm]{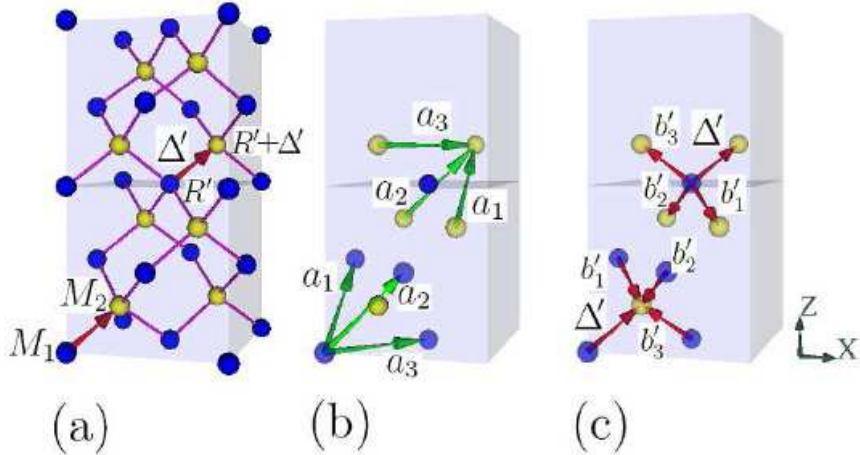}
\end{center}
\caption{Relevant vectors joining lattice points that are needed to discretize the 
displacement field in a zincblende lattice. All coordinates are expressed in the non-orthogonal 
basis spanned by the primitive vectors $a_{1}$, $a_{2}$, and $a_{3}$. (a) The basis of a unit
cell placed at $R'=(l,m,n)$ comprises one atom of mass $M_{1}$ with displacement vector
$u'_i(R';t)$ and one atom of mass $M_2$ and displacement vector $v'_i(R'+\Delta';t)$. (b)
Discrete gradients involving lattice points closest to $R'$ (resp. $R'+\Delta'$) are backward 
differences from $R'+\Delta'$ (resp., forward differences from $R'$) along the primitive 
directions: $D^-_{j}v'_{i}(R'+\Delta';t)$, (resp., $D^+_{j}u'_{i}(R';t)$), $i,j=1,2,3$. (c) 
The auxiliary vectors $b'_{i}$ satisfy $a'_{i}+b'_{i}=\Delta'$, $i=1,2,3$.}
\label{fig1}
\end{figure}
\end{center}

Once we have written the strain energy of a fcc crystal in the non-orthogonal basis
spanned by the primitive vectors, we can introduce our discrete model. We shall consider 
a fcc lattice with a two-atom basis. In equilibrium, atoms with mass $M_{1}$ will be placed 
at the lattice sites, so that their displacement vectors will depend on integer numbers and 
time: $u'_{i}= u'_{i}(l,m,n;t)$. In equilibrium, atoms with mass $M_{2}$ will be placed 
at the sites of a fcc lattice which, in the orthonormal basis $e_{i}$, is rigidly displaced by a 
vector $\Delta=(a/4,a/4,a/4)$ with respect to the first fcc lattice; see Fig.~\ref{fig1}. In 
terms of the non-orthogonal basis $a_{i}$, the vector $(a/4,a/4,a/4)$ becomes $\Delta'=
(1/4,1/4,1/4)$. 

We should define discrete differences of a displacement vector so that a discrete
difference become the corresponding partial derivative in the continuum limit. This
requirement can be satisfied in more than one way using different neighbors of a given
lattice point. We shall select only two neighbors of a lattice point for this purpose, using 
that the nearest neighbors of an atom with mass $M_{1}$ are atoms with mass $M_{2}$ and 
viceversa. Fig.~\ref{fig1} shows that each atom with mass $M_1$ (resp., $M_2$) is 
linked to its four nearest neighbors having mass $M_2$ (resp., $M_1$) by $\Delta'$, $b'_i=
\Delta'-a'_{i}$ (resp., $-\Delta',\, -b'_i$). Thus the nearest neighbors of an atom with 
displacement vector $v'_{i}(R'+\Delta';t)$ have displacement vectors $u'_{i}(R';t)$ and 
$u'_{i}(R'+a'_{j};t)$, with $j=1,2,3$, and the nearest neighbors of an atom with displacement 
vector $u'_{i}(R';t)$ have displacement vectors $v'_{i}(R'+\Delta';t)$ and $v'_{i}(R'+
\Delta'-a'_{j};t)$, with $j=1,2,3$. These facts motivate our definition of discrete differences 
of a displacement vector.

 Let us define the standard forward and backward difference operators along the
primitive directions as 
\begin{eqnarray} 
D^\pm_j f(R') = \pm [f(R'\pm a'_{j}) - f(R')].    \label{m2}
\end{eqnarray}
Then $\Omega_{ ij}^{(2)}(R'+\Delta';t)=D^+_{j}u'_{i}(R';t)$ is the discrete gradient of the 
displacement vector $v'_{i}(R'+\Delta';t)$ which involves lattice points closest to $R'+
\Delta'$, whereas $\Omega_{ij}^{(1)}(R';t)=D^-_{j}v'_{i}(R'+\Delta';t)$ is the discrete 
gradient of the displacement vector $u'_{i}(R';t)$  which involves lattice points closest to 
$R'$. The distortion tensor of our discrete model at the cell $R'$ could be defined as a 
weighted average of $g(\Omega_{ij}^{(1)})$ and $g(\Omega_{ij}^{(2)})$, in which $g(x)$ 
is a period-one periodic function such that $g(x)\sim x$ as $x\to 0$. For simplicity, we shall 
adopt equal weights in our definition:
\begin{eqnarray}
w_{i}^{(j)}(R';t) = {g\left(D^+_{j}u'_{i}(R';t)\right)+g\left(D^-_{j}v'_{i}(R'+
\Delta';t)\right)\over 2}. \label{m3} 
\end{eqnarray}
Obviously, this is reasonable for materials such as Si having a diamond structure, and also in 
the case of atoms of similar size for materials with zincblende structure. In the continuum 
limit $a\to 0$, the distortion tensor tends to the gradient of the displacement vector: 
\begin{eqnarray}
w_{i}^{(j)}\sim {\partial\tilde{u}'_{i}\over \partial x'_{j}}.  \label{m4}
\end{eqnarray}
In practice, the period-one function $g$ should be fitted to experimental or molecular 
dynamics data, such as the Peierls stress needed for a dislocation to move; see Fig.~1 of
Ref. \cite{CB05} for the variation of the Peierls stress as a function of the parameter 
controlling the shape of a piecewise linear function $g$. The positive potential energy of the 
crystal will therefore be
\begin{eqnarray}
V = {a^3\over 4}\sum_{l,m,n} {1\over 8} c'_{rspq} [g(D^+_s u'_r) + g(D^-_s v'_r)]\, 
[g(D^+_q u'_p) + g(D^-_q v'_p)].      \label{m5}
\end{eqnarray}
Here $a^3/4$ is the volume spanned by the three primitive vectors. 

The conservative equations of motion are   
\begin{eqnarray}
M_{1}\, {\partial^2 u'_i \over \partial t^2} = - \mathcal{T}^{-1}_{iq}
\mathcal{T}_{pq}^{-1} {\partial V \over \partial u_p'}. \label{m6} \\ 
M_{2}\, {\partial^2 v'_i \over \partial t^2} = - \mathcal{T}^{-1}_{iq}
\mathcal{T}_{pq}^{-1} {\partial V \over \partial v_p'}. \label{m7} 
\end{eqnarray}
Using Eq.\ (\ref{m5}), these equations become
\begin{eqnarray}
{4 M_{1}\over a^3}\, {\partial^2 u'_i \over \partial t^2}= {1\over 4}\, 
\mathcal{T}_{iq}^{-1} \mathcal{T}_{pq}^{-1} D_{j}^{-} \{ c'_{pjrs}\, g'(D^+_j u'_p)\, [g(
D_{s}^{-} v'_r) + g(D^+_s u'_r)]\},     \label{m8}\\
{4 M_{2}\over a^3}\, {\partial^2 v'_i \over \partial t^2}= {1\over 4}\,
\mathcal{T}^{-1}_{iq} \mathcal{T}_{pq}^{-1}  D^+_j \{c'_{pjrs}\, g'(D^-_j v'_p)\, [g(D^+_s u'_r) +
g(D^-_s v'_r)]\}.   \label{m9}
\end{eqnarray}
If $M_{1}=M_{2}$ (the case of Si), this system of equations is invariant under the 
symmetry: $u'_{i}(R')\leftrightarrow - v'_{i}(R'+\Delta')$. To obtain the continuum 
limit, we add Eqs.\ (\ref{m8}) and (\ref{m9}), take into account the continuum limit 
(\ref{m4}), and use Eqs.\ (\ref{fcc2}) to revert to the dimensional orthogonal 
coordinates. Then the resulting equations are those of anisotropic elasticity:
\begin{eqnarray}
\rho\, {\partial^2 \tilde{u}_i\over \partial t^2} &=& {\partial\over\partial 
x_{j}}\left( c_{ijrs}\, {\partial \tilde{u}_r\over\partial x_{s}}\right). 
\label{m10}
\end{eqnarray}
In this equation, the mass density $\rho$ is the sum of the masses in the primitive cell 
divided by the volume thereof: 
\begin{eqnarray}
\rho = {M_{1}+M_{2}\over a^3/4}. \label{m11}
\end{eqnarray}

Fluctuations and dissipation can be added to these equations in the same way as for the
models with one-atom basis. The derivation of these equations in non-orthogonal coordinates 
follows those in Sections \ref{sec:model} and \ref{dissipation}, but using the following 
energy instead of (\ref{B12}):
 \begin{eqnarray} 
{d\over dt}\,\sum_{l,m,n}\left[\sum_i {\rho\over 2}\left(\sum_{j}\mathcal{T
}_{ij}\dot{u}'_{j}\right)^2 + \sum_{rspq} {1\over 2} c'_{rspq} g(D^+_{s}u'_{r})\, 
g(D^+_{q}u'_{p}) \right]\nonumber\\
= \sum_{l,m,n} \sum_{i,j} \dot{u}'_i\, D^-_j
[\Sigma'_{ij}\, g'(D^+_j u'_i)]. \label{noc1}
\end{eqnarray} 
In particular, (\ref{B21}) also holds in non-orthogonal coordinates, which then yields the 
same formulas for the fluctuations as in orthogonal coordinates. For a zincblende structure,
the governing equations including dissipation and fluctuations are:
\begin{eqnarray} 
&& \rho c\,{\partial T\over\partial t}  + {T\over 2}\alpha'_{ij}\left(g'(D_{j}^+
u'_{i})\, D_{j}^+{\partial u'_i \over \partial t} + g'(D_{j}^-v'_{i})\, D_{j}^-{
\partial v'_i \over \partial t} \right)\nonumber\\
&&\quad\quad = D^-_i \left(\kappa'_{ij}D^+_{j}T + {\xi'_i
\over a}\right), \label{fd1}\\
&& \langle \xi'_i\rangle = 0,\nonumber\\
&& \langle \xi'_i(l,m,n;t) \xi'_j(l',m',n';t')\rangle 
= k_{B}T^2\, {\kappa'_{ij} + \kappa'_{ji}\over a}\, \delta_{ll'} \delta_{mm'} 
\delta_{nn'}\, \delta(t-t'),     \label{fd2}\\
&&{4 M_{1}\over a^3}\, {\partial^2 u'_i \over \partial t^2}= {1\over 2}\, 
\mathcal{T}_{iq}^{-1} \mathcal{T}_{pq}^{-1} D_{j}^{-} \{ [\sigma'_{pj}+
\Sigma'_{pj}+s'_{pj} ]\, g'(D^+_j u'_p)\},     \label{fd3}\\
&&{4 M_{2}\over a^3}\, {\partial^2 v'_i \over \partial t^2}= {1\over 2}\,
\mathcal{T}^{-1}_{iq} \mathcal{T}_{pq}^{-1}  D^+_j \{[\sigma'_{pj}+\Sigma'_{pj}
+s'_{pj} ]\, g'(D^-_j v'_p)\},   \label{fd4}\\
&& \langle s'_{ij}\rangle=0, \nonumber\\
&& \langle s'_{ij}(l,m,n;t) s'_{ab}(l',m',n';t') \rangle 
= k_{B}T {\eta'_{ijab} + \eta'_{abij}\over a^3} \delta_{ll'} \delta_{mm'} 
\delta_{nn'}\, \delta(t-t'),  \label{fd5}\\
&& \sigma'_{pj}= c'_{pjab}e'_{ab}- \alpha'_{pj}(T-T_{0}), \quad \Sigma'_{pj}=
\eta'_{pjab}\dot{e}'_{ab}, \label{fd6}
\end{eqnarray}
together with (\ref{em12}). In these equations, $e'_{ab}=(w_{a}^{(b)}+w_{b}^{(a)})/2$ is 
the symmetric part of the distortion tensor (\ref{m3}).

\subsection{Dislocations in Si and GaAs}
\label{sec:gaas}

\begin{center}
\begin{figure}[h]
\begin{center}
\includegraphics[width=8cm]{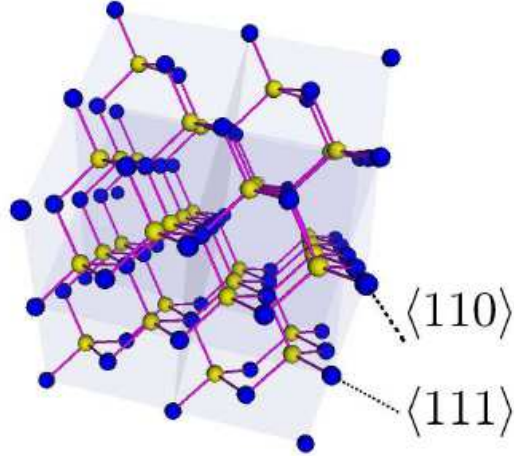}
\end{center}
\caption{Relevant directions in a fcc lattice with a two-atom basis. }
\label{fig2}
\end{figure}
\end{center}

Si and GaAs crystals are face-centred cubic with two atoms per 
lattice site, one at $(0,0,0)$ and the other one at $\Delta=(1/4,1/4,1/4)$. Both 
atoms are identical in Si (diamond structure), whereas they are different in GaAs: 
one atom is gallium and the other arsenic (zincblende structure). Each atom is 
tetrahedrally bonded to four nearest-neighbors, and the shortest lattice vector 
$\langle 110\rangle/2$ links a second neighbor pair, as shown in Figure \ref{fig2}. 
The covalent bond between two atoms is strongly localized and directional, 
and this feature strongly affects the characteristics of dislocations. In turn, 
dislocations influence both mechanical and electrical properties of these 
semiconductors \cite{hul01,hir82}. In this Section, we shall construct straight 
dislocations for Si and GaAs using their elastic stiffnesses and a method for calculating 
their strain field.

Perfect dislocations have Burgers vectors $\mathbf{b}=\langle 1 1 0\rangle/2$ (the same
as in the case of fcc lattices with a one-atom basis) and slip on the close packed $\lbrace 
1 1 1\rbrace$ planes having normal vector $\mathbf{n}$. Their dislocation line
vectors $\mathbf{\xi}$ usually lie along $\langle 1 1 0\rangle$ directions, forming 
$60^{\circ}$ (perfect $60^{\circ}$ dislocations) or $0^{\circ}$ (perfect screw 
dislocations) with respect to the Burgers vector. 

We will now construct a perfect screw and a perfect $60^{\circ}$
dislocation for GaAs lattice, both having Burgers vector
$\mathbf{b}=(1,0,1)/2$. Gliding dislocations having
this Burgers vector will leave behind a perfect crystal, since it
is a primitive vector of the lattice \cite{hul01}. For Si the same construction can be used.

The tensor of elastic constants is given by (\ref{e4}) in terms of the Voigt elastic 
stiffnesses $C_{ij}$ and the degree of anisotropy $H$ of (\ref{e5}).
We use the following values of the Voigt stiffnesses measured in units of $10^9$Pa at room 
temperature ($T=298$ K) \cite{hjort94}:
\begin{eqnarray}
C_{11}=165.6, &C_{12}=63.98,&C_{44}=79.51,\quad \textrm{for
Si;}\label{csi}\\
C_{11}=118.8,&C_{12}=53.8,&C_{44}=58.9,\quad \textrm{for
GaAs.}\label{cgaas}
\end{eqnarray}
Between $200$ K and $800$ K, these constants decrease linearly with
increasing temperature, so that $-C_{ij}^{-1}\, d C_{ij}/dT \approx
10^{-4}\textrm{K}^{-1}$, $i,j=1,2,3$, \cite{COTTAM,elassi,elasgaas}.
Such small corrections could be straightforwardly included in our
calculations, modifying minimally our results.

To calculate the elastic far field of any stationary straight dislocation, we shall follow
the method explained in Chapter 13 of Hirth and Lothe's book \cite{hir82}.
Firstly, we determine  an orthonormal coordinate system $e_1''$, $e_2''$, $e_3''$
with $e_3'' = -\xi$ parallel to the dislocation line and $e_2''=\mathbf{n}$ being the 
unitary vector normal to the glide plane. In terms of the new basis, the elastic displacement 
field $(u_1'',u_2'',u_3'')$ depends only on $x_1''$ and on $x_2''$.

Secondly, we calculate the elastic constants in the reference system $e_1''$, $e_2''$, $e_3''$:
\begin{eqnarray}
c_{ijkl}''=c_{ijkl} - H \sum_{n=1}^3 (S_{in}S_{jn}S_{kn}S_{ln}
-\delta_{in}\delta_{jn}\delta_{kn}\delta_{ln}). \label{m15}
\end{eqnarray}
Here the rows of the orthogonal matrix $S= (e_1'',e_2'',e_3'')^t$ are 
the coordinates of the $e_i''$'s in the old orthonormal basis $e_1$, $e_2$, $e_3$.
In the new reference system, the Burgers vector has coordinates 
$(\textrm{b}_1'',\textrm{b}_2'',\textrm{b}_3'')$. 

Thirdly, the displacement vector $(u_1'',u_2'',u_3'')$ is calculated as follows:
\begin{itemize}
\item Select three roots $p_1,p_2,p_3$ with positive imaginary
part out of each pair of complex conjugate roots of the polynomial
det$[a_{ik}(p)]=0$, $a_{ik}(p)=c_{i1k1}''+(c_{i1k2}''+c_{i2k1}'')p+
c''_{i2k2}p^2$.
\item For each $n=1,2,3$ find an eigenvector $A_k(n)$ associated to the
zero eigenvalue for the matrix $a_{ik}(p_n)$.
\item Solve Re$\sum_{n=1}^3 A_i(n)D(n)=\textrm{b}''_i$, and
Re$\sum_{n=1}^3 \sum_{k=1}^3 (c''_{i2k1}+c''_{i2k2}p_n)A_k(n)D(n) = 0$, in which
$i=1,2,3$, for the imaginary and real parts of $D(1)$,$D(2)$,$D(3)$.
\item For $k=1,2,3$, $u_k''=$ Re$[-{1\over 2\pi i}\sum_{n=1}^3
A_k(n) D(n)\ln (x_1''+p_n x_2'')]$.
\end{itemize}
Lastly, we can calculate the displacement vector $u'_k$ in the non-orthogonal basis $a_i$ 
from $u''_k$.

\begin{center}
\begin{figure}[h]
\begin{center}
\includegraphics[width=12cm]{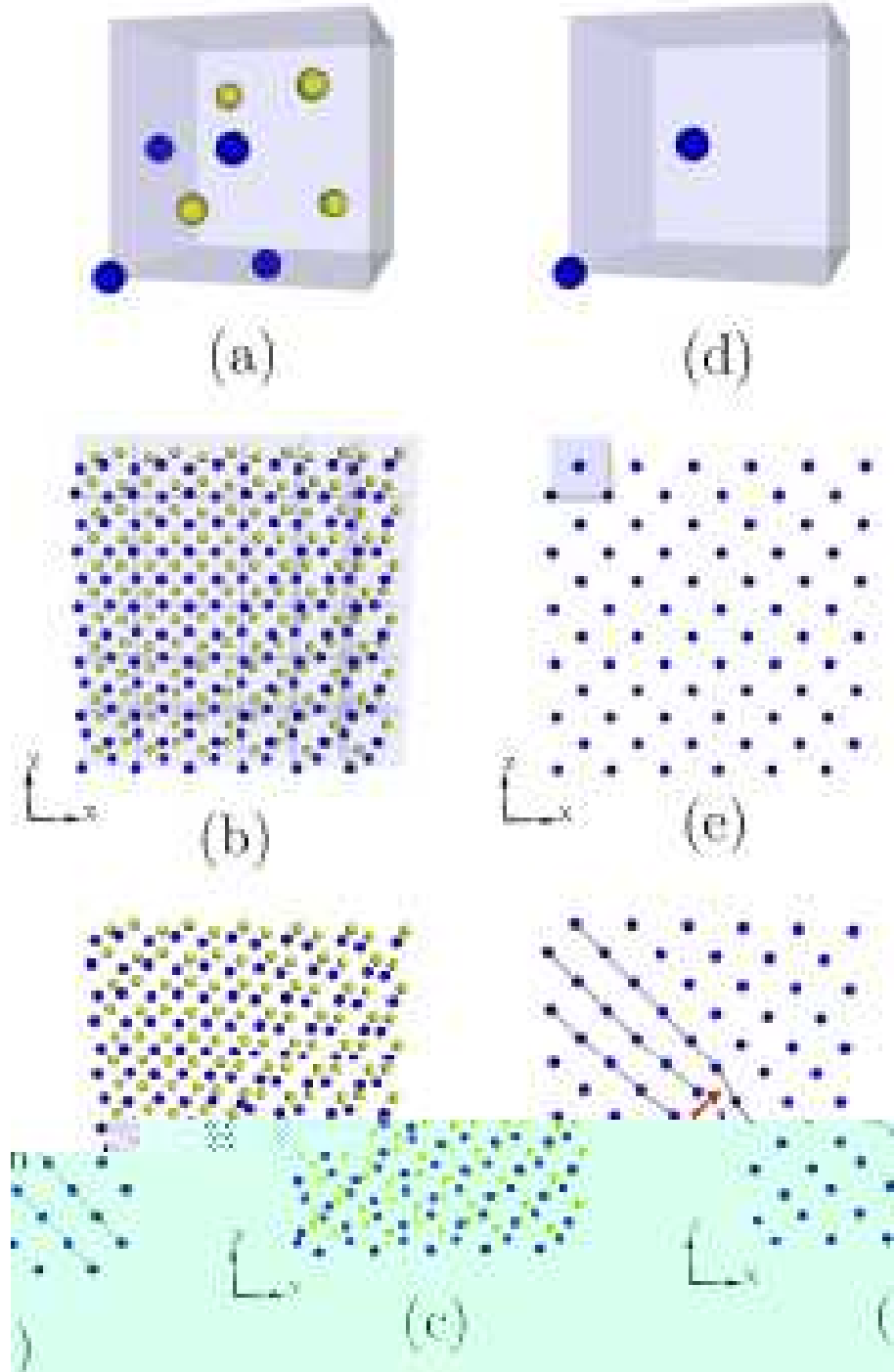}
\end{center}
\caption{Displacement field in a GaAs lattice created by a perfect
$60^{\circ}$ dislocation of Burgers vector
$\mathbf{b}=(1,0,1)/2$. (a) Reference cubic cell with its eight atoms.
(b) One layer of a perfect undistorted lattice. (c) The same layer distorted by a perfect 
$60^{\circ}$ dislocation. Panels (d), (e) and (f) correspond to (a), (b) and (c),
respectively but we have depicted only two atoms per reference cubic cell. }
\label{60dislocation}
\end{figure}
\end{center}

For the perfect $60^{\circ}$ dislocation, we have
\begin{eqnarray}
e_1''={1 \over \sqrt{6}}(1,1,2),\quad e_2''={1 \over
\sqrt{3}}(-1,-1,1),\quad e_3''={1 \over \sqrt{2}}(-1,1,0),
\end{eqnarray}
and the resulting dislocation is depicted in Fig. \ref{60dislocation}.

\begin{center}
\begin{figure}[h]
\begin{center}
\includegraphics[width=12cm]{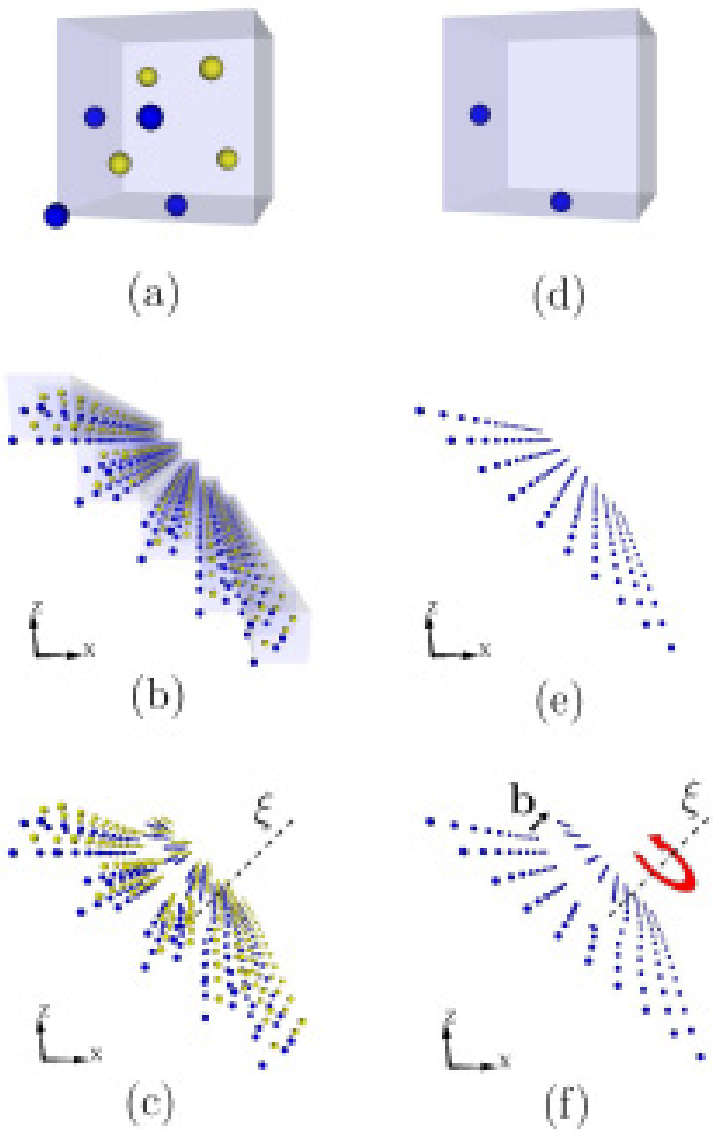}
\end{center}
\caption{Displacement field in a GaAs lattice created by a perfect
screw dislocation of Burgers vector $\mathbf{b}=(1,0,1)/2$. 
(a) Reference cubic cell with its eight atoms. (b) One layer of cubic cells normal to
the Burgers vector for a perfect undistorted lattice. (c) The same layer distorted by a perfect 
screw dislocation. Panels (d), (e) and (f) correspond to (a), (b) and (c),
respectively but we have depicted only two atoms per reference cubic cell. } \label{screw}
\end{figure}
\end{center}

For the pure screw dislocation, we have
\begin{eqnarray}
e_1''={1 \over \sqrt{6}}(-1,-2,1),\quad e_2''={1 \over
\sqrt{3}}(-1,1,1),\quad e_3''={1 \over \sqrt{2}}(-1,0,-1),
\end{eqnarray}
and the resulting dislocation can be observed in Fig. \ref{screw}.

\section{Conclusions} 
\label{sec:conclusions}
We have proposed discrete models describing defects in crystal structures whose
continuum limit is the standard linear anisotropic elasticity. The main
ingredients entering the models are the elastic stiffness constants of the
material and a dimensionless periodic function that restores the
translation invariance of the crystal (and together with the elastic constants
determines the dislocation size). For simple cubic crystals, their equations of motion
with conservative or damped dynamics (including fluctuations according as in Fluctuating
Hydrodynamics) are derived. For fcc and bcc metals, the primitive vectors 
along which the crystal is translationally invariant are not orthogonal. Similar discrete
models and equations of motion are found by writing the strain energy density and the 
equations of motion in non-orthogonal coordinates. In these later cases, we can determine
numerically stationary perfect edge and screw dislocations. We have also extended our discrete
models to the case of fcc lattices with a two-atom basis, which includes important applications
such as Si and GaAs crystals. For GaAs, we have calculated numerically two perfect
dislocations which may be used to calculate the structure and motion of similar dislocations
under stress as explained in Ref.~\cite{CB05}. Similarly to the case of the linear diatomic 
chain in which there are acoustic and optical branches of the dispersion relation, we expect 
that the dynamics of the discrete models with two-atom bases are richer than their continuum 
limits.

\section*{Acknowledgements}
This work has been supported by the Spanish Ministry of Education grants 
MAT2005-05730-C02-01 (LLB and IP) and MAT2005-05730-C02-02 (AC), and by the
Universidad Complutense grants Santander/UCM PR27/05-13939 and CM/UCM 910143 
(AC). I. Plans was financed by the Spanish Ministry of Education. He acknowledges Prof. 
Russel Caflisch and his Materials Modeling in Applied Mathematics group for their 
hospitality and fruitful discussions during a visit to UCLA and also Hyung-Jun Kim for 
his helpful explanations about the geometry of dislocations in Si.


\end{document}